\documentclass[letter,times,twocolumn,10pt]{article}
 \usepackage{usenix,epsfig,endnotes,microtype}
\pdfoutput=1
\usepackage[normalsize,bf]{caption}
\usepackage[compact]{titlesec}

\usepackage{amsmath, amssymb, amsfonts}
\usepackage{epsfig, graphicx, subfigure}
\usepackage{url}
\usepackage{algorithmic}
\usepackage{wrapfig}
\usepackage{times}
\usepackage{algorithm}
\usepackage{latexsym}
\usepackage{mdwlist}

\def\etal{et al. }

\author{%
{Ragib Hasan, Randal Burns}%
\vspace{1.6mm}\\
\fontsize{10}{10}\selectfont\itshape
Department of Computer Science, Johns Hopkins University\\
3400 N. Charles Street, Baltimore, MD 21218, United States\\
\{ragib,randal\}@cs.jhu.edu\\
}

\date{}
\title{The Life and Death of Unwanted Bits: Towards Proactive Waste
Data Management in Digital Ecosystems}

\begin{document}
\maketitle
    \abstract{Our everyday data processing activities create massive amounts of
data. Like physical waste and trash, unwanted and unused data also
pollutes the digital environment by degrading the performance and
capacity of storage systems and requiring costly disposal. In this paper, we
propose using the lessons from real life waste management in handling
waste data. We show the impact of waste data on the performance and
operational costs of our computing systems. To allow better waste data
management, we define a waste hierarchy for digital objects and
provide insights into how to identify and categorize waste data.
Finally, we introduce novel ways of reusing, reducing, and recycling
data and software to minimize the impact of data wastage.}
    \section{Introduction}
\label{sec:intro}

In real life, all human activities produce unwanted, unusable, or useless by-products. Such worthless objects are considered to be waste or trash. Waste products impact the environment and ecosystem by using up or polluting resources, degrading performance of physical processes, and requiring expensive cleanup. To deal with waste in the ecosystem, various waste management techniques have been developed over the years \cite{pongracz2004evolving,pongracz2004re}. These techniques aim at reducing the production of waste, repurposing the waste or its components, and efficient disposal of waste.

In today's world, we are increasingly living virtual lives -- creating, processing and consuming data in the form of digital objects. A computing system is similar to a real life ecosystem. In a digital ecosystem, data and applications that consume and produce data interact and use the physical hardware components and resources. Like real life ecosystems, a digital ecosystem has finite resources such as storage, compute cycles, and network bandwidth. Also, consumer applications share and compete for these resources.  No matter how much ``illusion'' of infinite resources various abstractions provide, in reality resources in a digital ecosystem are not infinite. Storing, transferring, and disposing of data consume these resources. However, all data in a given system are not equally important or useful, and often a significant amount of data in a system can be unwanted or unused content. Such waste data consumes resources but provide no value to the corresponding digital ecosystem. Unless we introduce responsible waste data management practices, such waste data will misuse resources and make a significant impact on the digital ecosystem.

What happens to these unwanted bits? Typical approaches to managing waste data include compression and/or deletion of such unwanted data. However, these processes come at a price -- disposal of waste data consumes resources in the form of energy used to delete data, tying up compute cycles, blocking I/O, etc. Disposal via deletion also causes degradation of performance and reduces the lifetime of storage components (such as Flash storage). We need better waste management techniques to handle unwanted data. In this paper, we argue about the need to examine waste data in a systematic manner. We posit that successful real life waste management techniques can be effectively adapted to handle waste data. The contributions of this paper are as follows:
\vspace{-5pt}
\begin{enumerate*}
\item We present a definition of waste data in digital ecosystems.
\item We show the impact of waste data on reduction and degradation of system capacity.
\item We introduce a hierarchy for waste data management techniques.
\item We advocate the need for an integrated approach for managing waste data and discuss how well known real life waste management principles can be adapted for this.
\end{enumerate*} 
\vspace{-5pt}
The rest of the paper is organized as follows: Section~\ref{sec:wastedef} presents a definition of waste data. We discuss the impact of waste data in Section~\ref{sec:impact}. We introduce a hierarchy for waste data management and explore the use of different real life waste management principles for managing waste data in Section~\ref{sec:hierarchy}. Finally, we discuss related work in Section~\ref{sec:related} and conclude in Section~\ref{sec:conclusion}.

	\section{Defining Waste Data}
\label{sec:wastedef}

In real life, the definition of waste is somewhat subjective, as what is waste to one system can be considered valuable resources by another system. Various authorities and agencies have defined waste in different ways. For example, the Basel convention and the European Union define waste as something that is or will be discarded by the holder \cite{75eec,basel}. The Organization for Economic Cooperation and Development (OECD) defines waste as materials that are by-products of regular processing, which have no use to the creator and which are disposed of \cite{oecd}. Pongr{\'a}cz \etal \cite{pongracz2004evolving,pongracz2004re} provided a definition of waste based on their classification of objects -- an object is considered to be waste if it is unintentionally created, or the user has used up the object, or the object's quality has degraded, or the object is unwanted by the user. 

Similarly, providing the definition of waste data in a computing system is difficult. Informally, a data object or software can be considered to be waste data by a user if it has no utility for the user in the given context. To provide a formal definition of waste data, we leverage the definition of physical waste given by Pongr{\'a}cz \etal \cite{pongracz2004evolving,pongracz2004re}. In particular, we use Pongr{\'a}cz et al.'s classification scheme to define waste data as data belonging to any one or more of the following categories:
\vspace{-7pt}
\begin{itemize*}
\item \textbf{Unintentional data.~} Data unintentionally created, as a side effect or by-product of a process, with no purpose.
\item \textbf{Used data.~} Good data that has served its purpose and is no longer useful to the user.
\item \textbf{Degraded data.~} Data that has degraded in quality such that it is no longer useful to the user.
\item \textbf{Unwanted data.~} Data that was never useful to the user.
\end{itemize*}
\vspace{-7pt}
Next, we discuss each of these waste data categories with examples.\\

\vspace{-7pt}
\noindent\textbf{Unintentional data.~} Almost all data processing applications generate unintentional by-products. We define a data object to be a by-product if it is not included in the final set of data objects produced by the application. For example, the goal of LaTeX compilation is to generate output .pdf or .ps files from source text and images. However, when LaTeX is executed, it automatically creates a number of temporary data objects and files that assist in compilation. In the context of LaTeX compilation, these files (such as .aux, .bbl, .log) can be considered to be unintentional by-products that assist in the production of the final data product (e.g., .ps or .pdf). \\

\vspace{-7pt}
\noindent\textbf{Used data.~} In most cases, input data is no longer useful to the user once computations have been performed over it. For example, an aggregation operation can use data from many sensors. The sensor readings may be useful to the user performing the aggregation operation only until the computation is over. After that, the input data may become useless, and hence considered to be waste data.\\

\vspace{-7pt}
\noindent\textbf{Degraded data.~} When data gets corrupted, it can become unusable, and therefore be considered as waste data by a user. Also, when other changes make data or software obsolete, it can be considered to have degraded and therefore marked as waste data by the user. For example, newer software releases can make old versions of the software and the related files obsolete.\\

\vspace{-7pt}
\noindent\textbf{Unwanted data.~} This class of waste data includes data that may or may not be of high quality, but is not relevant to the user at all to begin with. For example, software documentation in an unknown language can be of good quality but still be irrelevant to a non-speaker.
	\section{Impact of Waste Data}
\label{sec:impact}

In the physical reality, waste has adverse impact on the environment of the ecosystem. The presence of waste pollutes the ecosystem, causing economic, social, and operational impacts. We argue that in the same manner, waste data affects a computing system by consuming resources without providing value, and by degrading system performance and components. Storage, processing, and transfer of data require the use of system resources such as disk space, cpu cycles, and network or I/O bandwidth. Disposing of the waste data by deletion also impacts the life of storage devices and incurs energy and time overheads. \\

\vspace{-7pt}
\noindent\textbf{Storage Consumption.~} Waste data consumes a lot of storage space. For example, creating and editing a small text file in  \textit{vi} causes a temporary swap file to be created. To illustrate the amount of temporary waste data created by source code compilation, we compiled Openssl 1.0.0a on a Linux workstation. Compilation of Openssl produces about 13.6 MB of target binary code. However, it also produces about 44.5 MB of temporary object code that is not part of the installation. From the viewpoint of the user, these temporary object files produced during compilation are waste data. Such unwanted data consumes a lot of space and needs to be deleted.

The overall amount of unused and dead data in a given system is not small. We wrote a Perl script to determine the percentage of files that have never been accessed since last modification. We ran the script on three different platforms -- an Apple MacBook used as a personal laptop, a Ubuntu Linux desktop, and a student lab server running Fedora Linux. The results are show on Table~\ref{table:1}. 
\begin{table}[tdp]
\begin{center}
\begin{tabular}{|c|c|c|c|}
\hline 
\textbf{Platform} & \textbf{MacBook} &  \textbf{Desktop} & \textbf{Server} \\
\hline
\textbf{\% of files} & 20.6 & 47.4 & 57.1 \\
\hline
\textbf{\% of used space} & 98.5 & 38.1 &  99.5 \\	 
\hline
\end{tabular}
\end{center}
\caption{\textbf{Analysis of files in a MacBook, a desktop workstation, and a student lab server. In all cases, a large number (20.6\%--57.1\%) of files have never been accessed since last modification.}}
\label{table:1}
\end{table}%
In all three cases, a large fraction of files in the system have never been accessed since last modification, reflecting the results from previous work in the area \cite{douceur99large,satyanarayanan1981study}. In terms of space usage, these files amounted from 38\% to as high as 99\% of the total used space on the machines. This shows that the amount of waste data in a system is quite significant. \\

\vspace{-7pt}
\noindent\textbf{Reducing Device Lifetime.~} Disposal of waste data via deletion can impact the lifetime of storage devices. For example, Flash based storage devices typically have a maximum number of write cycles. Multi-Level Cell (MLC) flash devices support a maximum of 1,000--10,000 write/erase cycles per cell while Single-Level Cell (SLC) flash devices support up to 100,000 write-erase cycles per cell \cite{wang10reuse}. Waste or by-product data brings no value, but uses up flash storage write cycles, reducing the lifetime of such storage devices. As flash-based solid state storage becomes popular, especially in mobile devices, we need to ensure that waste data write / erase cycles do not impact the lifetime of such storage.\\ 

\vspace{-7pt}
\noindent\textbf{Performance Degradation.~} Presence of unwanted waste data can degrade system performance. For example, in a file system, the extra storage space consumed by waste data may cause unnecessary fragmentation and use up available inodes. If waste data can be identified and not stored by the system, we can reduce the load and fragmentation greatly. For example, Table~\ref{table:1} shows that 98.5\% of the total space used by files in the laptop was actually consumed by files that were never accessed since last modification. By storing these files separate from the frequently accessed files, we can vastly improve system performance. Deletion also takes up CPU cycles and consumes energy -- a fact which is significant in low-powered mobile devices.

	\section{Managing Waste Data}
\label{sec:hierarchy}
How do we deal with waste data? Storage and deletion of waste data is costly in terms of energy and space consumed. Therefore, we need effective strategies to handle waste data. To provide a guideline for waste data management, we turned to the techniques used in real life waste management. We argue in this paper that these lessons from real life waste management are equally effective in managing waste data in digital environments. We start our discussion by presenting a hierarchy of waste data management methods. Then we discuss some specific approaches application designers and system architects can adopt to minimize the impact of waste data.
\begin{figure}[tbp]
\centering
\includegraphics[width=3.1in]{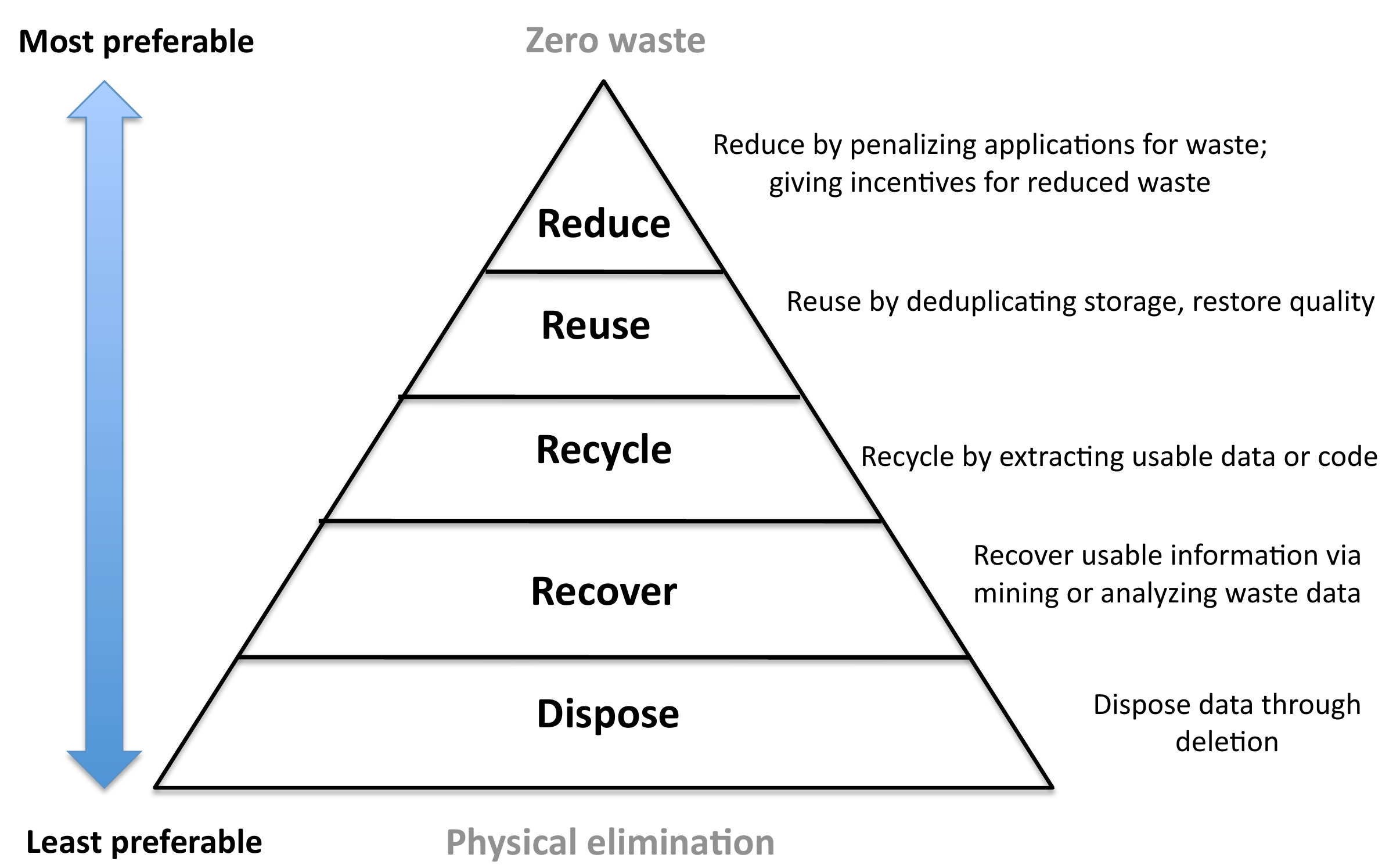}
\caption{\textbf{The Waste Data Management Hierarchy. Processes at the top are more preferable.}}
\label{fig:hierarchy}
\end{figure}
\subsection{Waste data hierarchy}
In dealing with waste in the natural environment, a waste hierarchy is widely used to classify and organize waste management schemes according to their usefulness and impact \cite{wilson1996stick}. We propose adapting the waste hierarchy from real life waste management to develop guidelines for choosing waste data management schemes. Besides the ``three R's'' (\textit{reduce, reuse, recycle}), we use the additional steps of recovery and disposal in our scheme, leveraging the five-step waste management hierarchy described in \cite{75eec}. We show the waste data hierarchy in Figure~\ref{fig:hierarchy}, and describe the steps below:\\

\vspace{-7pt}
\noindent\textbf{Reduce.~} At the top of the waste data hierarchy is \textit{reduction of waste}, which refers to reducing the amount of waste data generated in the system. We opine that this is the most favorable option, since less waste will cause the least overhead on the system. Applications should be designed with waste-reduction in mind and only store the desired output data in the disk. In-memory caching of temporary values and content-based addressing can help reduce the amount of waste data produced by applications. Operating systems and file systems can provide incentives to applications that produce less waste data and punishments to those that produce a lot of waste data (we discuss this later in this section).
\\

\vspace{-7pt}
\noindent\textbf{Reuse.~} In the next layer, we have  \textit{reuse of waste}, which refers to reusing the waste data for other purposes. Schemes that can be classified as data reuse include data deduplication \cite{zhu08dedup}, where the content of waste data objects can be used by the deduplication scheme to achieve better compression ratios. Another example is the reuse of translation memories in machine translation, where the information from one translation session can be used to enrich global translation capabilities. Google's Translation Toolkit already allows this type of data reuse, and it has been used successfully for translation of English Wikipedia articles into African and South Asian languages \cite{googletrans}. For degraded data, regeneration \cite{zadok2004reducing} or restoration \cite{schroeder2010understanding} can be used to recover data quality.
%

\noindent\textbf{Recycle.~} Slightly less preferable than reuse is to \textit{recycle waste data}, where data objects can be broken up and used for different objects. While it is difficult to fathom what it means to recycle application specific data for other purposes, we can definitely recycle waste containing application code. When an obsolete software package is going to be removed, we can extract the usable components from it and use them for other applications. 
\\

\vspace{-7pt}
\noindent\textbf{Recover.~} Sometimes, the waste data cannot be recycled or reused. A possibility of still gaining some utility from such data is to \textit{recover} information. For example, used log files can be anonymized and shared or analyzed for getting high-level views. Obsolete data can also be mined to gather patterns about historical trends. 
\\

\vspace{-7pt}
\noindent\textbf{Dispose.~} At the bottom of the hierarchy sits schemes for \textit{Disposal of data}, through deletion. This is costly in terms of time and energy spent deleting data objects. So, we opine that deletion should be the absolute last recourse in managing waste data.

Above the hierarchy is the ideal state of \textit{zero waste}, where careful system design results in production of no waste data. Below this hierarchy of schemes for waste data management lies another approach not shown in the hierarchy -- \textit{physical elimination}. Sometimes, the five schemes may not be enough or feasible for managing waste data. For example, the data may be stored in physically immutable media, and hence not subject to any of the above schemes. Also, security issues and regulations may require physical elimination of the storage media. This can be achieved by incinerating, degaussing, or destroying the storage media. However, this has the worst impact on the natural environment as any such disposal would impact the physical ecosystem.

Next, we discuss some specific strategies and best practices for waste data management.

\vspace{-3pt}
\subsection{Some schemes for managing waste data}
Taking a cue from successful real life waste management strategies, we propose several schemes for managing waste data in digital ecosystems. For this, we leverage the concepts of waste hierarchy and extended producer responsibility \cite{lifset1993take}. \\

\vspace{-7pt}
\noindent\textbf{Digital landfills.~} A digital landfill is the equivalent of real-life landfills, where unwanted data can be disposed of without additional cost associated with deletion. For this, we propose using a semi-volatile storage device. Such a storage device would store data, but gradually unwanted data objects will fade automatically and the storage space can be reclaimed. This type of device can be implemented on a volatile storage medium using a least-recently-used (LRU) scheme, where data which has not been used recently is allowed to fade, while more frequently used data is refreshed. 

\noindent\textbf{Waste penalties for applications.~} A \textit{waste penalty} can be imposed on applications that create large amounts of waste data. For example, the operating system can penalize an application that creates a lot of temporary files by reducing its I/O bandwidth or schedule it to receive fewer CPU cycles. This gives applications incentives to act responsibly in creating waste data. This concept is equivalent to the \textit{Pay-as-You-Throw} scheme and the \textit{polluter-pays principle} used in real life waste management \cite{payt}.\\

\vspace{-7pt}
\noindent\textbf{Extensive system-wide Deduplication and Micro-modular software.~} A big problem with recycling old or unwanted software is that software libraries are not usually written to allow extraction of small amounts of code. Shared dynamic link libraries do allow code sharing among multiple applications \cite{mili2002reusing}. However, they do not allow removal of unused routines to retain only the routines that are used. When recycling a library, it is therefore not possible to extract usable routines from it. To allow recycling old code, we propose breaking up software code libraries into micro-modules in the level of individual routines or algorithms, which can be extracted from the library when recycling old code.
	\section{Related Work}
\label{sec:related}
While researchers have explored different storage management issues, the systematic management of waste data has received little interest. Information lifecycle management (ILM) has been used by the storage industry to determine optimal management of data objects throughout their lifecycle \cite{reiner2004information}. A major challenge in ILM is to  design valuation schemes to determine the importance of information. Chen presented such a scheme based on file access patterns \cite{chen2005information}. We can use such schemes to identify waste data. Zadok \etal \cite{zadok2004reducing} advocated for the need to reduce storage consumption through application of regeneration and smart space reclamation policies, in order to increase device lifetimes and available storage. Researchers have also analyzed existing systems to identify typical usage patterns. An early work by Satyanarayanan \cite{satyanarayanan1981study} introduced the notion of functional lifetime (f-lifetime) for files, defined as the difference between a file's age and the time since its last access. Files with lower f-lifetimes are less useful, since the gap between their creation/last modification times and last access times are short. In a study of file systems, Douceur \etal \cite{douceur99large} showed that 44\% of the files in the studied systems had an f-lifetime of zero (the percentage was higher, at 67\%, for technical support systems), indicating that these files have not been accessed at all since last modification. This agrees with our findings presented in Section~\ref{sec:impact}. Vogels found that file lifetimes are often quite short -- almost 80\% files are actually deleted within 4 seconds \cite{vogels99sosp}. The very short lifetime indicates that the usefulness of these files ends quickly. Deleting these files is costly, and application designers should rethink their I/O to prevent the creation of such waste data. Finally, researchers have developed techniques such as software refactoring and reuse \cite{mili2002reusing}, and data deduplication \cite{zhu08dedup}, which can be applied in different stages of the waste data hierarchy to reduce the impact of waste data.

    \section{Conclusion}
\label{sec:conclusion}

For many years, the abundance of storage space and decreasing storage costs have allowed us to ignore the adverse impact of waste data on our digital ecosystems. But as we start dealing with massive quantities of data, we need to manage waste data in order to reduce overheads and energy costs, and improve efficiency. In this paper, we defined the waste data problem and proposed using techniques from real life waste management to minimize the impact of waste data on our computing environment. Our waste data management hierarchy can be used to determine the preferable option in dealing with waste data in different applications. We also advocated the adoption of responsible application behavior and best practices in reducing the impact of waste data. We posit that software engineering techniques as well as hardware architectures will need to be adapted with waste data minimization, management, and recycling in mind, in order to build a efficient and sustainable digital ecosystem.

\section*{Acknowledgements}
This work was supported by the National Science Foundation under Grant \#0937060 to the Computing Research Association for the CIFellows Project. 
\bibliographystyle{abbrv} 
\bibliography{paper}
\end{document}